\documentclass[%
superscriptaddress,amsmath,amssymb,preprint,floatfix,longbibliography,noeprint, 
]{revtex4-1}
\usepackage{graphicx,xcolor}
\usepackage{mathrsfs,dsfont,mathtools}
\usepackage{dcolumn}
\usepackage{bm}
\usepackage[colorlinks, citecolor=blue, linkcolor=blue, urlcolor=blue]{hyperref}
\usepackage[english]{babel}
\usepackage[babel,kerning=true,spacing=true]{microtype}
\usepackage{braket}

\usepackage{threeparttable}

\definecolor{DarkRed}{RGB}{100,0,0}

\begin{document}

\title{Charge density waves and electronic properties of superconducting kagome metals}

\author{Hengxin Tan}
\author{Yizhou Liu}
\affiliation{Department of Condensed Matter Physics,
Weizmann Institute of Science,
Rehovot 7610001, Israel}
\author{Ziqiang Wang}
\affiliation{Department of Physics, Boston College, Chestnut Hill, Massachusetts 02467, USA}
\author{Binghai Yan}
\email{binghai.yan@weizmann.ac.il}
\affiliation{Department of Condensed Matter Physics,
Weizmann Institute of Science,
Rehovot 7610001, Israel}

\date{\today}

\begin{abstract}
Kagome metals $A$V$_3$Sb$_5$ ($A=$ K, Rb, and Cs) exhibit intriguing superconductivity below $0.9 \sim 2.5 $ K, a charge density wave (CDW) transition around $80\sim 100 $ K, and $\mathbb{Z}_{2}$ topological surface states. The nature of the CDW phase and its relation to superconductivity remains elusive. In this work, we investigate the electronic and structural properties of CDW by first-principles calculations. We reveal an \textit{inverse} Star of David deformation as the $2\times2\times2$ CDW ground state of the kagome lattice. 
The kagome lattice shows softening breathing-phonon modes, indicating the structural instability. However, electrons play an essential role in the CDW transition via Fermi surface nesting and van Hove singularity. The inverse Star of David structure agrees with recent experiments by scanning tunneling microscopy (STM). The CDW phase inherits the nontrivial $\mathbb{Z}_{2}$-type topological band structure. Further, we find that the electron-phonon coupling is too weak to account for the superconductivity $T_c$ in all three materials. It implies the existence of unconventional pairing of these kagome metals. Our results provide essential knowledge toward understanding the superconductivity and topology in kagome metals.
\end{abstract}

\maketitle

The kagome lattice constitutes uniformly tiled triangles and hexagons in a plane. Related insulating materials are intensively studied for intriguing phenomena like spin liquid \cite{shores2005}, geometrical frustration\cite{Ramirez1994}, charge density wave (CDW)\cite{Isakov2006}, and unconventional superconductivity of doped Mott insulators\cite{Ko2009}. For the more recent interest of topology, several magnetic kagome metals were found to be magnetic Weyl semimetals\cite{Yang2017,morali2019fermi,liu2019magnetic} and exhibit a large anomalous Hall effect\cite{Nakatsuji2015,Nayak2016,Liu2018,ye2018massive}. Very recently, nonmagnetic kagome metals $A$V$_3$Sb$_5$ ($A=$ K, Rb, and Cs)\cite{Ortiz2019materials} were reported to exhibit both CDW and superconductivity, and $\mathbb{Z}_{2}$-type nontrivial band topology\cite{Ortiz2020CVS,Ortiz2020KVS}, which sparked immediate interest in these materials\cite{Yin2021RVS,Yang2020,Yu2021,Wang2020}. 

The $A$V$_3$Sb$_5$ compounds exhibit a uniform kagome lattice of V atoms at room temperature.  Calculations and angle-resolved photoemission spectroscopy\cite{Ortiz2020CVS} found the $\mathbb{Z}_{2}$ type topology in their band structure.
These materials go through a CDW transition as cooling down to about $80\sim 100 $ K. From the CDW phases, they exhibit superconductivity with T$_c = 0.9 \sim 2.5 $ K (see Table \ref{Table_SC} for a summary). In CsV$_3$Sb$_5$, very recent experiments reveal possible nodal superconductivity \cite{Zhao2021CVSnodal} and the competition between superconductivity and CDW\cite{Chen2021CVS}. For the CDW phase, X-ray diffraction\cite{Ortiz2020CVS,Ortiz2020KVS} and scanning tunneling microscopy (STM)\cite{Jiang2020STM,Zhao2021CVS,Liang2021,Chen2021} revealed the formation of a $2 \times 2$ superlattice, for which Ref. \onlinecite{Jiang2020STM} assumed a Star of David (SD) distortion in the kagome plane and the proximity of the Fermi surface to the van Hove singularity. However, solid knowledge of the lattice and electronic properties for the CDW state is missing to understand the superconductivity and topology.

In this work, we have investigated the structural and electronic properties of the CDW phase by first-principles density-functional theory \cite{SM} calculations. We find that the CDW transition is related to breathing phonon modes of the kagome lattice and electronically mediated by the Fermi surface instability. The resultant $2\times 2$ CDW favors an \textit{inverse} Star of David (ISD) deformation rather than the SD energetically for all three compounds. The STM images simulated for the ISD phase agree with the recent experiment. The CDW distortion modulates the Fermi surface but preserves the $\mathbb{Z}_{2}$ topology of the pristine phase. We find the electron-phonon coupling too weak to interpret the superconductivity for three materials. It may indicate the existence of unconventional superconductivity pairing.

\begin{figure}
\centering
\includegraphics[width=\linewidth]{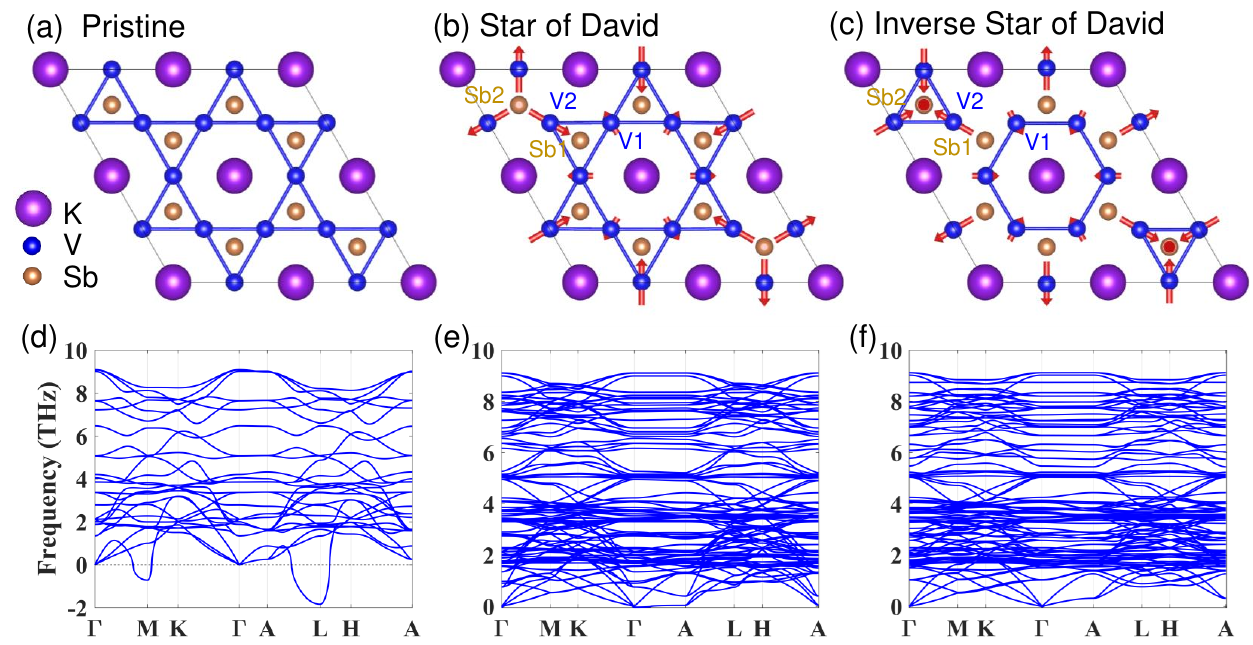}
\caption{
Crystal structures of KV$_3$Sb$_5$ in the pristine phase, the Star of David 2$\times$2 CDW phase, and the inverse Star of David phase. Corresponding phonon dispersions are shown in the lower panels. A 2$\times$2 supercell is shown in (a). Imaginary (negative) phonon frequency in (d) corresponds to the breathing mode of the kagome lattice. Such a breathing-instability is related to CDW distortions. The breathing-out and breathing-in lead to two different structures in (b) and (c), where the red arrows indicate the lattice distortion due to the breathing mode. In CDW phases, V1 and V2 represent inequivalent vanadium sites of the kagome lattice, and Sb1 and Sb2 are inequivalent sites of the honeycomb lattice.}
\label{fig1_str_phonon}
\end{figure}




\begin{figure}
\centering
\includegraphics[width=\linewidth]{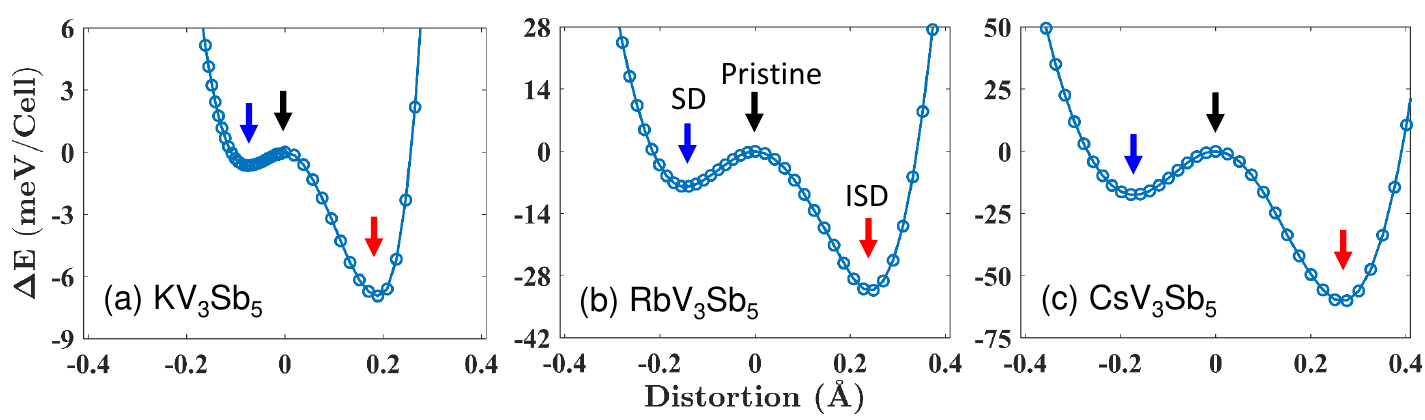}
\caption{\label{fig2_tot_energy}
Total energy profiles for the three compounds. The distortion stands for the amplitude of the breathing phonon mode where the negative values are used for Star of David (SD) in order to distinguish from inverse Star of David (ISD). The $\Delta$E stands for the relative total energy with respect to the pristine phase per supercell (36 atoms). Spin-orbit coupling is considered.}
\end{figure}

\begin{figure}
\centering
\includegraphics[width=\linewidth]{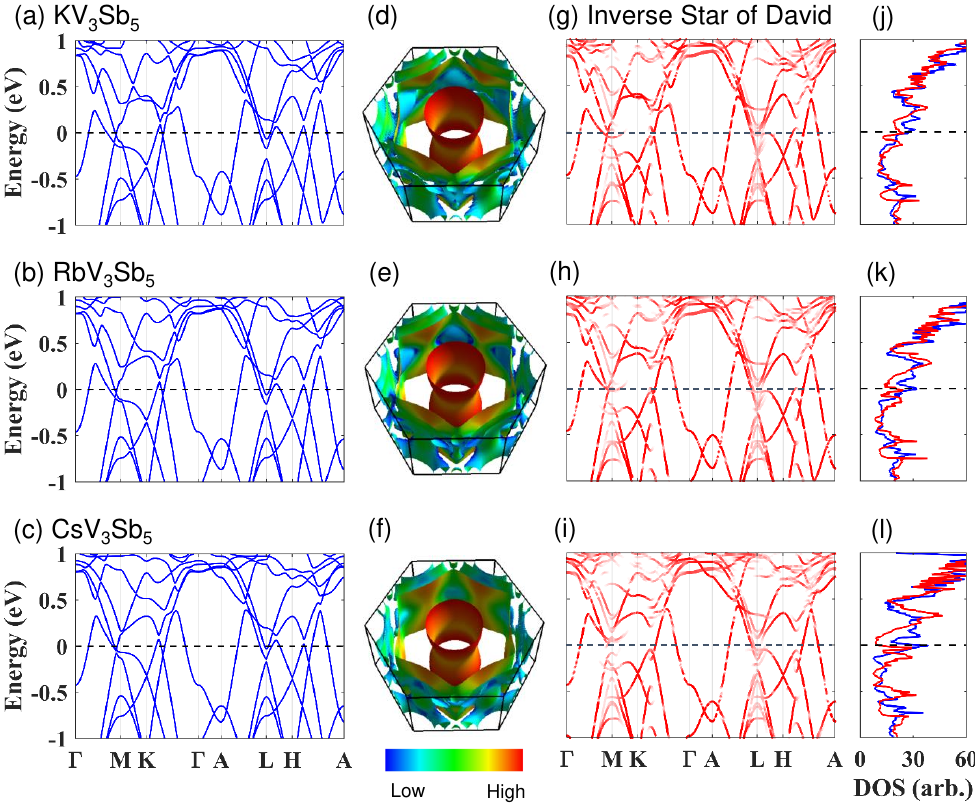}
\caption{\label{fig3_band}
Electronic structures. (a-c) show the band structure for the pristine phase. (d-f) show the Fermi surface for the pristine phase, where the color stands for the Fermi velocity. (g-i) show the unfolded band structure of the inverse Star of David phase where the color represents the $1\times 1$ spectrum weights. (j-l) show the total DOS of the pristine (blue) and inverse Star of David (red) phases. Spin-orbit coupling is included.}
\end{figure}

\begin{figure}
\centering
\includegraphics[width=\linewidth]{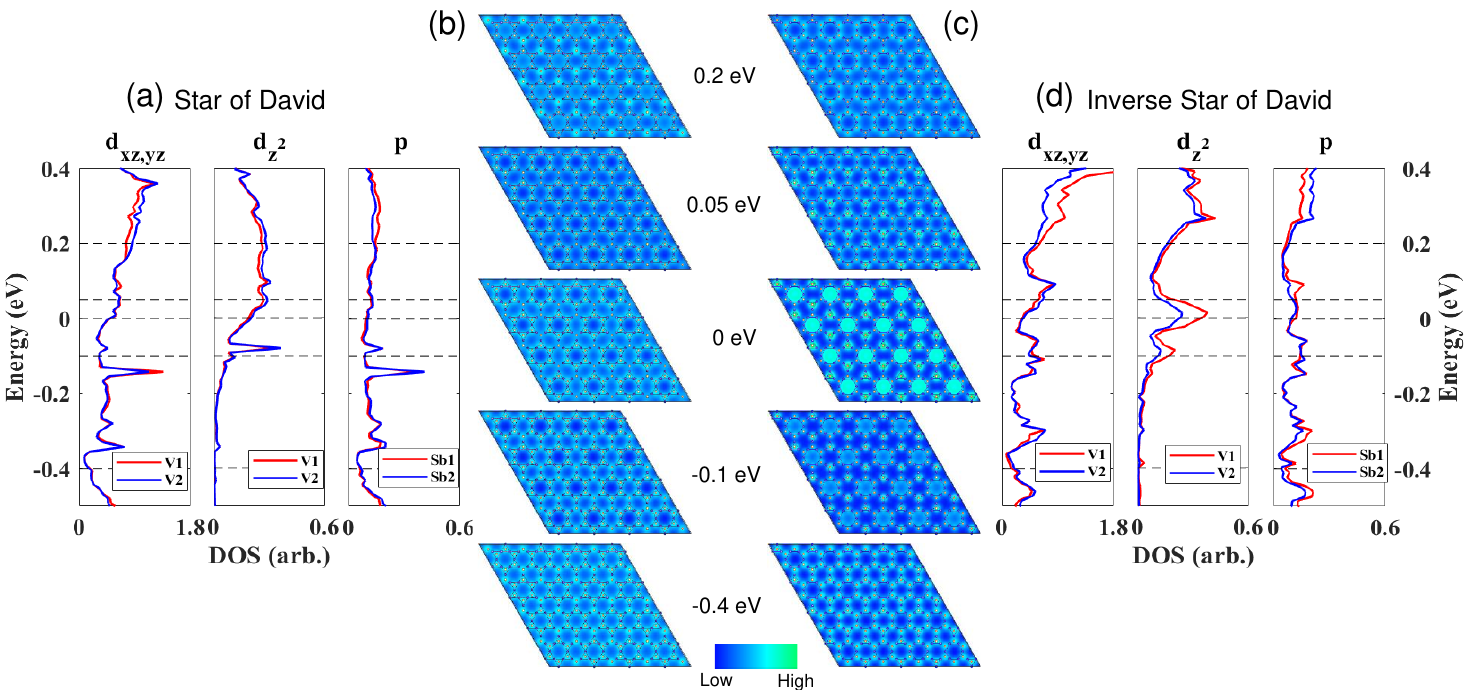}
\caption{\label{fig4_STM}
Surface orbital-resolved DOS and simulated STM images for the $2\times 2$ CDW phases of KV$_3$Sb$_5$: (a) and (b) for the Star of David structure, (c) and (d) for the inverse Star of David structure. The DOS of V-$d$ and Sb-$p$ states are projected. The V1, V2, Sb1 and Sb2 in (a) and (d) are labeled in Fig. \ref{fig1_str_phonon} (b) and (c) and superposed on the images for reference.}
\end{figure}

\textit{CDW deformation --}
The pristine phase of $A$V$_3$Sb$_5$ crystallizes in the hexagonal structure with the space group of $P6/mmm$ (No.191), as shown in Fig. \ref{fig1_str_phonon}(a). The V atoms form the kagome layer, with Sb occupying V hexagon centers. Two additional honeycomb layers of Sb atoms sandwich the kagome lattice. Three materials are nonmagnetic in our calculations, consistent with experiments \cite{Ortiz2019materials,Kenney2021}.

The pristine phases exhibit structural instability. We take KV$_3$Sb$_5$ as an example. Its phonon band structure in Fig. \ref{fig1_str_phonon}(d) shows softening acoustic phonon modes at the Brillouin zone boundary near $M$ and $L$ points, indicating the strong instability. 
Extended to the $2\times 2$ supercell, the $M$-point soft mode corresponds to a breathing phonon ($A_{1g}$ symmetry) of V atoms in the kagome lattice [Figs. \ref{fig1_str_phonon}(b) and (c)]. Breathing-in and out lead to two different structures. The SD structure forms when V1 (V2) atoms move away (toward) the center, which is similar to the known CDW in 1T TaS$_2$\cite{Wilson1975}. The ISD structure forms by an inverse deformation to the SD. Both SD and ISD exhibit an in-plane 2$\times$2 modulation of the structure. The superlattice shares the same space group of $P6/mmm$ as the pristine phase but has two inequivalent sites (i.e., V1 and V2 in Figs. \ref{fig1_str_phonon}) in the kagome plane.
In addition, the $L$-point soft mode [Fig.\ref{fig1_str_phonon}(d)] indicates the existence of a $2\times 2 \times 2 $ reconstruction. We will first investigate the 2$\times$2 CDW and then discuss the inter-layer stacking pattern.

The optimized SD and ISD structures are locally stable for all three compounds, where the breathing deformation reduces the total energy. The imaginary frequencies disappear in their phonon spectra except for a faint soft feature in the SD case (see more details in the Supplementary Material (SM)\cite{SM}). It should be noted that structure optimizations are independent from phonon calculations. We find that the SD phase is higher in energy than the ISD phase for all three materials. As shown in Fig. \ref{fig2_tot_energy}, from K, to Rb and Cs, the ISD phase is more energetically favored, and the lattice displacement also becomes larger and larger. Therefore, the ISD phase is more likely the ground state structure than SD in the $2\times 2$ CDW. In addition, the calculated diffraction pattern of ISD agrees better with experiments\cite{Ortiz2020CVS} than that of SD\cite{SM}. The ISD phase will be verified further by simulated STM images in the following.

The CDW is induced by the Peierls instability related to the Fermi surface nesting and phonon softening\cite{peierls1996quantum,Kohn1959}. The $q$ vector of the soft phonon at $M$ coincides with the Fermi nesting vector between neighboring van Hove singularities in the Fermi surface of $A$V$_3$Sb$_5$ (see Fig.\ref{fig3_band}). Consequently, the CDW deformation strongly decreases the density of states (DOS) at the Fermi energy by suppressing the van Hove singularities. Related Fermi surface nesting, CDW, and their competition with superconductivity are extensively discussed for the kagome lattice model in literature\cite{Wen2010,Yu2012,Kiesel2013,Wang2013}. 
Additionally,  the phonon-softening at $M$ and $L$ can be suppressed if we artificially increase the \textit{electron} temperature in DFT calculations \cite{SM}. Therefore, the electronic instability may play an essential role to drive the CDW transition, compared to phonons.

Above energetics and structures of two CDW phases are studied for bulk materials. When forming a surface with different charging or atomic-terminations, the ISD or SD structure may deviate on the surface. To examine the CDW stability on the surface, we have performed surface calculations using a slab model. Recent experiments\cite{Jiang2020STM,Zhao2021CVS,Chen2021} observed both $A$ and Sb terminations on cleavage surfaces. We find that the ISD phase is still as stable as the bulk phase for the $A$-termination. However, the ISD deformation becomes weaker (closer to the pristine case) on the Sb-terminated surface than in the bulk. The SD phase exhibits a similar trend. Nevertheless, the ISD structure is still more energetically favored than the SD one. The surface atomic relaxation indicates the instability of the $2\times 2$ CDW on the Sb-terminated surface. We note that recent STMs\cite{Chen2021,Zhao2021CVS} observed a 4$\times$1 pattern when the samples are cooled down further after the 2$\times$2 CDW transition. The emerging 4$\times$1 pattern may be due to electron correlations related to above surface instability and electron-phonon interaction.

Further, the very recent STM measurement\cite{Liang2021} revealed a three-dimensional CDW, i.e., the $2\times 2\times 2$ CDW, for CsV$_3$Sb$_5$. This coincides with the inter-layer instability characterized by the soft phonon mode at the $L$ (0.5, 0.5, 0.5) point in the phonon spectrum [Fig.\ref{fig1_str_phonon}(d)]. Our calculations confirm that the $\pi$ CDW-phase-shift between neighboring layers reduces slightly the binding energy by $\sim$ 10 meV per $2\times 2$ layer for CsV$_3$Sb$_5$ \cite{SM}. In the $2\times 2\times 2$ CDW, the SD phase is no longer meta-stable and relaxes to the ISD structure spontaneously. Therefore, the ISD deformation is still preferred. We note that the K and Rb compounds exhibit similar 3D CDW to the Cs one in calculations. Additionally, the $2\times 2\times 2$ CDW exhibits nearly the same electronic structure as the $2\times 2\times 1$ one, because of the weak inter-layer interaction. For simplicity of analysis, we focus on the $2\times 2\times 1$ (i.e., $2\times 2$) CDW phase in the following discussions.


\textit{Band structure and Fermi surface --}
The band structures of the pristine phase are shown in Fig. \ref{fig3_band}(a)-(c) for three materials, which are similar to the previous report. We should note that the Fermi energy (charge neutral point) in our work is about 0.1 eV higher than that of Ref.\cite{Ortiz2019materials}. Such a minor derivation lifts the Fermi energy above the van Hove singularity point at $M$ and leads to a slightly different Fermi surface (FS) near the Brillouin zone boundary [Fig. \ref{fig3_band}(d)-(f)]. The FSs exhibit a strong 2D feature and include the cylinder-like FS (Sb-$p_z$ states) centered around $\Gamma$, and the large hexagonal FS (V-$d$ states). There are also much smaller FSs (V-$d$ states) near $M/L$ and $K/H$. Additionally, one can also observe the 2D feature in the weak $\Gamma-A$ dispersion in the phonon spectrum [Fig.\ref{fig1_str_phonon}(d)].

The 2$\times$2 ISD CDW distortion modifies the FSs. We unfold the band structure of the ISD phase to the pristine Brillouin Zone in Figs. \ref{fig3_band}(g)-(i). These bands near the Brillouin Zone boundary (e.g., $M$ and $K$) are strongly modified, including the large hexagonal FS and small FSs. It is not surprising that the breathing distortion changes mainly the V-$d$ driven states and especially suppresses the van Hove singularities slightly below the Fermi energy at $M$. Consequently, CDW reduces the total density of states (DOS) near the Fermi energy [Fig. \ref{fig3_band}(j)-(l)], which is also consistent with recent STM spectra\cite{Jiang2020STM,Chen2021}. The Cs compound exhibits stronger FS and DOS modifications than the K and Rb compounds, because the lattice distortion is the largest for Cs (see Fig.\ref{fig2_tot_energy}).

Different from V-$d$ driven FSs, the cylinder-like FS from Sb-$p_z$ is marginally affected by the CDW deformation, as shown in Fig.\ref{fig3_band}. Because the breathing distortion mainly involves the V-kagome lattice, the CDW and electron-lattice interaction\cite{SM} are band-selective. This rationalizes the STM quasiparticle interference patterns in Refs. \onlinecite{Zhao2021CVS,Liang2021}, where the Sb-$p_z$ band around $\Gamma$ is not affected by the CDW (either $2 \times 2$ or $4 \times 1$). 

The band structure of the pristine phase displays nontrivial band inversion\cite{Ortiz2020CVS,Ortiz2020KVS}. For the CDW phases, we calculate the $\mathbb{Z}_{2}$ topological invariants\cite{Fu2007} and find that both SD and ISD phases are still nontrivial topological metals (see Ref.\cite{SM} for details).


\textit{Simulated STM images --}
STM images can examine the electronic and lattice structures of SD/ISD CDW in experiments. Therefore, we have simulated STM images for two CDW phases for all three materials. We also take the K compound as an example here since the Rb and Cs compounds share similar features\cite{SM}. We have calculated the local density of states (LDOS) in the plane that is $\sim$3 \AA~ above the Sb-terminated surface (in a slab model\cite{SM}) to simulate the STM dI/dV mapping. For a given bias $V_b$, the LDOS is summed in the energy range of $V_b \pm$ 25 meV. Figures \ref{fig4_STM}(a)\&(d) show the surface DOS contributed by V1, V2, Sb1 and Sb2 atoms [labeled in Fig. \ref{fig1_str_phonon}(b) and (c)], 
to rationalize the bias dependence of STM images.

The most characteristic feature near the Fermi energy is that the ISD structure exhibits large, bright $2\times 2$ spots by the V1-hexagon centers. In contrast, the SD structure exhibits bright circles around the V1-hexagon, with the hexagon center being dark. For ISD, the V1-hexagon centers turn quickly dark as moving away from the zero energy, which agrees with the recent STM for KV$_3$Sb$_5$ in Ref.\cite{Jiang2020STM}. Also supported by the lower total energy, we conclude that the ground state of the $2\times 2 $ CDW phase corresponds to the ISD structure for these three compounds. The high LDOS at V1-hexagon centers is caused by the large peak of V1-$d_{z^2}$ and small V1-$d_{xz,yz}$ in DOS [Fig.\ref{fig4_STM}(d)]. In contrast, no such a V1-$d_{z^2}$ peak appears for the SD surface [Fig.\ref{fig4_STM}(a)].

At $V_b=0.05$ eV, the V1-hexagon brightness dramatically decreases due to the decreasing DOS of V-$d_{z^2}$ on the ISD surface. At the same time, the V-$d_{xz,yz}$ DOS increases, where the V2-$d_{xz,yz}$ contribution is slightly larger than the V1-$d_{xz,yz}$ one. Thus, small V2-triangles turn brighter. For the SD surface, the STM image at $V_b=0.05$ eV remains almost the same as that at zero because the relative amplitudes of DOS between V1 and V2 are nearly unchanged. For $V_b=0.2$ eV, the ISD surface exhibits bright circles because V1 is larger in DOS than V2. In contrast, the SD surface shows small, bright spots at V2-triangles, caused by the relatively larger DOS of V2 than that of V1.

At $V_b=-0.1$ eV, the ISD surface looks less-contrast in the STM image than the zero bias case due to the decreasing V-$d_{z^2}$ and increasing V-$d_{xz,yz}$ in DOS. For the SD surface, all the atomic sites show similar brightness due to the small DOS peaks of V-$d$ and Sb-$p$ near $-$0.1 eV. At $V_b=-0.4$ eV, both ISD and SD surfaces show a 1$\times$1 honeycomb pattern. The honeycomb feature reflects the top Sb layer (Sb1 and Sb2), despite that V is comparable to Sb in DOS.

\textit{Superconductivity --}
$A$V$_3$Sb$_5$ exhibits superconductivity from the CDW normal state. The most recent STM on CsV$_3$Sb$_5$ reveals a superconducting gap $\Delta$  $\sim 0.5$ meV\cite{Chen2021}, which puts the superconductor in the strong-coupling regime with 2$\Delta/k_BT_c \sim 5$. A natural question is whether the superconductivity is driven by the electron-phonon coupling in Bardeen-Cooper-Schrieffer (BCS) scenario \cite{BCS1957}. Therefore, we have calculated the electron-phonon coupling for the ISD phase and estimated the $T_c$ based on McMillan's formula.
Table \ref{Table_SC} lists the calculated $T_c$ for ISD phases of three materials. The deformation potentials ($D$) are one order of magnitude smaller than the known BCS superconductor MgB$_2$ \cite{An2001MgB2}. The phonon-mediated attractive interaction($\lambda$) is smaller than 1, indicating that three materials are in the weakly-coupled regime. It is in contradiction with the experimental indication from the superconducting gap mentioned above. The smallness of $\lambda$ for CsV$_3$Sb$_5$ is mainly due to the low DOS ($N_F$) at the Fermi energy. The calculated $T_c$ values are much smaller than the corresponding experimental values. Especially, phonons alone cannot account for the experimentally determined strong coupling superconductivity in CsV$_3$Sb$_5$.
It indicates the unconventional mechanism of the superconductivity observed in these materials. 
In addition, we expect that the DOS reduction at the Fermi energy due to the CDW will lead to a Tc reduction. In this sense, the CDW and superconductivity are competing which is an interesting question for future study.

\begin{table}
\centering
\caption{\label{Table_SC} Estimation of the $T_c$ due to electron-phonon coupling. We list the CDW transition temperature $T^{\text{exp.}}_{\text{CDW}}$, experimental superconducting temperature $T_c^{\text{exp.}}$ and experimental Debye temperature $\theta^{\text{exp.}}_D$ in unit of K \cite{Ortiz2020KVS,Ortiz2020CVS,Yin2021RVS}. Parameters and results: 
DOS $N_F$ per superlattice at the Fermi energy (eV$^{-1}$), deformation potential $D$ (eV/\AA), phonon mediated electron-electron attractive interaction strength $\lambda$, calculated Debye temperature $\theta_D$, and calculated $T_c$ for the ISD phase. We choose the screened-repulsive interaction ($\mu^*$) as 0.12 \cite{An2001MgB2, Si2013Graphene}. More details are in SM \cite{SM}.}
\begin{threeparttable}
\begin{tabular}{c|cccccc|cc}\hline\hline
$\quad$ & $T^{\text{exp.}}_{\text{CDW}}$ & $N_F$  & $D$  & $\lambda$ & $\theta_D$ & $\theta^{\text{exp.}}_D$ & $T_c$  & $T^{\text{exp.}}_c$  \\\hline
KV$_3$Sb$_5$ & 78 & 11.6 & 0.47 & 0.38 & 154 & 141 & 0.22 & 0.93\\
RbV$_3$Sb$_5$ & 102 & 9.3 & 0.50 & 0.32 & 176 & 169 & 0.05 & 0.92 \\
CsV$_3$Sb$_5$ & 94 & 5.2 & 0.69 & 0.25 & 142 & -- & 0.0008 & 2.5  \\\hline\hline
\end{tabular}
\end{threeparttable}
\end{table}

In summary, we have studied the lattice and electronic structures of the CDW phases of three kagome metals $A$V$_3$Sb$_5$. The 2$\times$2 CDW (also the $2\times 2\times 2$ one) phase exhibits an inverse Star of David deformation. Corresponding STM images simulated agree with recent experiments. The nontrivial $\mathbb{Z}_{2}$ nature remains in the CDW band structure. We have evaluated the superconducting temperatures induced by electron-phonon coupling and found they are much smaller than  experimental values, implying possible unconventional superconductivity in these materials. 

\textbf{Acknowledgements.} We thank H. Chen for inspiring discussions and A. Perara for help in calculations. B.Y. acknowledges the financial support by the Willner Family Leadership Institute for the Weizmann Institute of Science, the Benoziyo Endowment Fund for the Advancement of Science,  Ruth and Herman Albert Scholars Program for New Scientists, the European Research Council (ERC) under the European Union's Horizon 2020 research and innovation programme (Grant No. 815869) and and ISF MAFAT Quantum Science and Technology (2074/19). Z.W.is supported by the U.S. Department of Energy, Basic Energy Sciences Grant No.DE-FG02-99ER45747.

\bibliography{references}

\end{document}